\documentclass[11pt]{article}

\usepackage[preprint]{acl}

\usepackage{times}
\usepackage{latexsym}
\usepackage{amsmath}
\usepackage{amssymb}
\usepackage{booktabs}
\usepackage{tabularx}

\usepackage[T1]{fontenc}

\usepackage[utf8]{inputenc}

\usepackage{microtype}

\usepackage{inconsolata}

\usepackage{graphicx}

\title{Latent Terms: Dense Retrievers Contain Trivially Extractable BM25-ready Zipfian Vocabularies}

\author{
 \textbf{Benjamin Clavié\textsuperscript{1,2}},
 \textbf{Sean Lee\textsuperscript{1}},
\\
\textbf{Aamir Shakir\textsuperscript{1}},
 \textbf{Makoto P. Kato\textsuperscript{2,3}}
\\
 \textsuperscript{1} Mixedbread AI, \\
 \textsuperscript{2} National Institute of Informatics (NII),
 \textsuperscript{3} University of Tsukuba
\\
 \small{
   \textbf{Correspondence:} \href{mailto:ben@mixedbread.com}{ben@mixedbread.com}
 }
}

\begin{document}
\maketitle
\newcommand{\method}{\emph{Latent Terms}}
\newcommand{\saebm}{SAE-BM25}
\newcommand{\beir}{BEIR-13}
\newcommand{\msmarco}{MS MARCO}
\newcommand{\gte}{GTE-ModernColBERT}

\begin{abstract}
We propose \textbf{\method{}}, a method revealing that models trained for dense retrieval, whether single- or multi-vector, learn representations that can trivially be decomposed into retrieval-ready sparse features.
When trained on frozen retrievers, Sparse Autoencoders without any retrieval-specific adjustments extract a latent vocabulary with approximately Zipfian collection statistics, directly suitable for classical sparse retrieval scoring via BM25.
This approach enables sparse retrieval while requiring no learned expansion objective or sparse retrieval supervision whatsoever, and can be readily applied to any dense retriever. \method{} is able to match or outperform single-vector scoring methods from its own base model as well as comparable SPLADE variants. In addition, it substantially outperforms its base model on LIMIT, a task specifically designed to highlight the failures of single-vector retrieval.
Overall, our results highlight that neural retrievers contain more expressive and indexable structure than their default scoring functions expose, but that other methods can nonetheless be leveraged.
\end{abstract}

\section{Introduction}

Neural information retrieval is deeply tied to representation learning. A retrieval model, typically built on a pre-trained language model backbone, is trained to produce representations that can be searched through a particular scoring interface~\cite{neuralir}. In practice, neural retrievers are often categorized by the representations they expose at inference time and by the operators used to score them. Dense single-vector retrievers encode queries and documents into one vector each and score them with dot product or cosine similarity~\cite{dpr}. Late-interaction, or dense multi-vector, retrievers expose sets of token-level vectors and score them with operations such as MaxSim~\cite{colbert}. Learned sparse retrievers such as SPLADE expose sparse vocabulary weights that can be indexed and searched efficiently~\cite{splade}. 


Meanwhile, Sparse Autoencoders (SAEs), models trained to map a model activation into a higher-dimensional sparse code and then reconstruct the original activation from that code, have become a widely used tool for analyzing the internal representations of neural networks~\cite{SAE1,topksae}.


In this work, we ask whether the interface exposed by a retriever captures all of the retrieval-relevant structure learned by the model. Recent work has shown that single-vector retrievers can sometimes be adapted into strong multi-vector retrievers~\cite{jacolbertv2.5,gtemoderncolbert}, suggesting that trained retrievers may encode useful retrieval structure beyond what their default scoring interface exposes. We study a complementary question: do dense retrievers also contain sparse, indexable structure, even when they are not trained to produce sparse representations?

Specifically, we hypothesize that SAEs could recover such structure from dense retrievers, by converting a model's representations into a "quasi-lexical" latent vocabulary. 
To test this hypothesis, we introduce \textbf{\method{}}. Given a frozen retriever, \method{} encodes queries and documents, projects their final-layer token representations through an SAE, and applies BM25~\cite{bm25} directly over the resulting sparse activations, treating activated feature indices as vocabulary terms and transformed activation magnitudes as term weights.

Importantly, \method{} does not train a sparse retriever with retrieval supervision, nor use any normally-required sparse training methods such as learned expansion objectives~\cite{splade}, hard negatives~\cite{hardnegs}, or sparsity regularization, with FLOPs regularization being the most common and alternatives remaining an active area of research~\cite{flopsalt}. Instead, the SAE is trained only with the standard SAE reconstruction objective over web text extracted from FineWeb-Edu~\cite{finewebedu}. All sparsity comes from the SAE itself, while ranking is performed by a classical BM25 scorer over the resulting features.

We apply \method{} to multiple dense retrievers with varying original retrieval performance. Despite its simplicity, \method{} consistently extracts strong sparse retrieval performance from all evaluated frozen backbones: it matches comparable SPLADE variants\footnote{Comparable models defined as competitive models developed around the same time period.} and outperforms the base model's single-vector cosine similarity approach on both single-vector backbones tested. The gains are especially pronounced on benchmarks designed to expose limitations of single-vector models, further suggesting that \method{} can leverage relevant structure that is present in the model but inaccessible through its single-vector scoring interface. 



We then show that SAE features learned from retrieval models form a latent vocabulary whose collection statistics resemble those of natural-language terms, providing BM25 with meaningful document-frequency statistics.
Qualitative analysis supports the view that SAEs extract a meaningful vocabulary, which contains a mixture of lexical as well as both narrow and broad semantic units, combining sparse indexability with a vocabulary induced from the neural retriever's internal representation.

Overall, our results suggest a different view of neural retrieval models. A model's default scoring function is not necessarily the only useful way to access its retrieval knowledge. Dense retrievers can contain sparse, expressive, and indexable structure that their inference interface does not expose, and this structure can be recovered with a reconstruction-trained SAE and classical sparse IR methods such as BM25.

\textbf{Contributions} In summary, our contributions are:
We (i) introduce \method{}, a simple method for converting frozen retriever activations into BM25-searchable sparse representations using reconstruction-trained SAEs; (ii) show that these latent vocabularies support strong sparse retrieval without sparse retrieval supervision; and (iii)  propose an analysis of why the method works by showing that the generated vocabulary has term-like collection statistics and a mix of meaningful semantic and lexical units.

\section{Background}
\label{sec:background}

\subsection{Sparse Autoencoders}
\label{sec:background-sae}

Sparse Autoencoders (SAEs) are shallow neural networks trained to represent a dense activation
$h \in \mathbb{R}^{d}$ using a higher-dimensional sparse code
$z \in \mathbb{R}_{\geq 0}^{m}$, where typically $m \gg d$.
They are built on an encoder-decoder architecture, made up of an encoder $f_{\mathrm{enc}}$ and decoder $f_{\mathrm{dec}}$:
\begin{equation}
    z = f_{\mathrm{enc}}(h), \qquad \hat{h} = f_{\mathrm{dec}}(z),
\end{equation}
Trained jointly with an objective comprising a reconstruction term, encouraging information preservation, and a sparsity penalty so each input activates only a small subset of latent features:
\begin{equation}
    \mathcal{L}_{\mathrm{SAE}}(h)
    =
    \| h - \hat{h} \|_2^2
    +
    \lambda \| z \|_1 .
\end{equation}

SAEs have become a common tool in neural network, and specifically language-model interpretability. In the latter, the activations on which the SAE is trained are individual token-level activations. This approach is used to decompose dense neural activations into features that are more localized and interpretable than individual coordinates of the original representation~\cite{SAE1}, thus facilitating the process of interpreting otherwise ``black-boxed'' neural activations~\cite{gemmascope,anthro}. Indeed, rather than interpreting activations through potentially polysemantic individual dimensions, SAEs aim to learn a basis in which different latent dimensions can map to a specific pattern or concept~\cite{SAEinterp}.



\subsection{Okapi BM25}
\label{sec:background-bm25}

Okapi Best Match 25, more frequently referred to as just BM25~\cite{bm25}, is a ubiquitous method in classical information retrieval which remains surprisingly competitive against modern neural methods, especially with proper per-dataset parameter tuning~\cite{bm25optim}. Given a query $Q$ and a document $D$, BM25 scores $D$ by summing the contributions of query terms that occur in the document:
\begin{equation}
\begin{aligned}
    \operatorname{BM25}(Q,D)
    &=
    \sum_{t \in Q}
    \operatorname{IDF}(t)
    \frac{f(t,D)(k_1 + 1)}
    {f(t,D) + k_1 K_D},
    \\
    K_D
    &=
    1 - b + b \frac{|D|}{\operatorname{avgdl}} .
\end{aligned}
\end{equation}
Here, $f(t,D)$ is the frequency of term $t$ in document $D$, $|D|$ is the length of $D$, $\operatorname{avgdl}$ is the average document length in the collection, $k_1$ controls term-frequency saturation and $b$ controls document-length normalization. The inverse document frequency term is commonly defined as
\begin{equation}
    \operatorname{IDF}(t)
    =
    \log
    \frac{
        N - n(t) + 0.5
    }{
        n(t) + 0.5
    },
\end{equation}
where $N$ is the total number of documents and $n(t)$ is the number of documents containing term $t$.

Traditionally, BM25 is used directly on textual inputs, with various levels of pre-processing, functioning as a bag-of-words method defined over lexical terms. Its scoring combines inverse document frequency, term-frequency saturation, and document-length normalization.
However, its underlying assumptions are not inherently lexical: BM25 can in principle be applied to any set of sparse features with meaningful collection frequencies, magnitudes, and lengths. 



\subsection{Learned Sparse Retrieval}

Learned sparse retrieval encompasses a family of neural methods that
preserve the efficiency of classical lexical retrieval while leveraging language models to improve its representation. Broadly, the general recipe is to keep sparse,
vocabulary-indexed representations, but have them be learned or enhanced by a model rather than directly derived from the surface form of the input text.

This taken many forms throughout the years, with early instantiations addressing queries and documents separately. Initially, work focused on the document-side: DeepCT~\cite{deepct} and DeepImpact~\cite{deepimpact} developed contextualized methods to learn sparse document representations, while Doc2Query approaches~\cite{doc2query,doct5query} focused on vocabulary expansion to mitigate the document/query vocabulary mismatch problem. uniCOIL~\cite{unicoil} further expanded on these methods by attempting to reconcile them, learning scalar weights over lexical terms with optional vocabulary expansion.


Following this early work, SPLADE~\cite{splade} proposed handling weighting and expansion jointly within a single end-to-end model, leveraging the language modeling abilities of a pre-trained model such as BERT~\cite{bert}. Given an input sequence $x$, SPLADE reuses the language modeling head of its base encoder to project each contextualized token representation $h_i$ onto the encoder vocabulary $V$, before aggregating these projections into a single sparse vector $w(x) \in \mathbb{R}^{|V|}_{\geq 0}$:
\begin{equation}
    w_j(x)
    =
    \max_{i \in 1..|x|}
    \log\!\left(
        1 + \operatorname{ReLU}\!\left(
            \operatorname{MLM}(h_i)_j
        \right)
    \right) .
\end{equation}
with the log-ReLU transformation ensuring non-negative vocabulary weights and a final pooling operation allowing each vocabulary item to be activated by the most relevant input position. The resulting sparse vector can therefore contain both observed terms and expansion terms predicted by the language model. Scoring is then defined as sparse vector matches, commonly expressed with an inner product:
\begin{equation}
    s(q,d)
    =
    \langle w(q), w(d) \rangle
    =
    \sum_{j \in V} w_j(q) w_j(d).
\end{equation}
As the majority of coordinates are zero, these representations can be indexed and searched with efficient indexing methods such as inverted indexes, while still benefiting from contextualized neural term weighting and expansion.

Achieving competitive retrieval with SPLADE-style models, however, requires considerably more than simply applying a masked language-modeling head. In addition to the common complexities of retrieval training, such as mined hard negatives or knowledge distillation techniques~\cite{spladev3}, SPLADE's performance can be sensitive to factors other model families are robust to, such as the tokenization method~\cite{faire}, and requires explicit sparsity regularization during training~\cite{splade,spladepp}.

\subsection{Other SAE-Based Retrieval Work}

CL-SR~\cite{clsr} proposed the use of SAEs on the task of reconstructing the final, single-vector representations of a dense retriever. In doing so, they demonstrated that not only do the extracted features provide a degree of interpretability, and showed that the resulting latent features could be scored in a SPLADE-like manner, using an inner product to perform retrieval, which they dubbed a form of \textbf{C}oncept-\textbf{L}evel \textbf{S}parse \textbf{R}etrieval. However, while a promising avenue, its retrieval performance is substantially degraded compared to that of its original single-vector retriever and is built on a fully in-domain setting, with in-domain queries and the same corpus used for training the SAE and evaluating retrieval downstream. Furthermore, CL-SR does not explore the use of SAEs on token-level representations, and instead instead focusing on final, single-vector representations.

Concurrent work such as SPLARE~\cite{splare} and BM25-V~\cite{bm25v} have also both proposed different ways of leveraging SAEs as vocabularies. BM25-V~\cite{bm25v} proposes the BM25 over SAE-generated features from a CLIP~\cite{clip}-like vision encoder, but restricts its exploration to the use of such method as a high-recall, low-ranking-quality first stage retriever which requires a second stage using the model's normal scoring function instead. Meanwhile, SPLARE uses an SAE-generated vocabulary over a frozen LLM. This vocabulary is then used as the basis for full retrieval training, employing a SPLADE-like training pipeline and achieving moderate but consistent improvements over the same training pipeline using the original model vocabulary instead~\cite{splare}. 


\section{\method{}: BM25 over SAE Features Extracted from Dense Retrievers}
\label{sec:method}

At a high level, \method{} takes a frozen dense retriever $R$, trains an SAE on the activations $R$ produces over unlabeled text, and uses the resulting sparse latent code as a vocabulary on which BM25 is applied. This approach relies on three broad steps: training the SAE, constructing latent representation, and BM25 scoring over the latent vocabulary.

\subsection{Training SAEs on Frozen Retrievers}
\label{sec:method-training}

Let $R$ be a frozen dense retriever that maps an input sequence
$x = (x_1, \ldots, x_{|x|})$ to a set of contextualized token representations
\begin{equation}
    R(x) = (h_1, \ldots, h_{|x|}), \qquad h_i \in \mathbb{R}^d ,
\end{equation}
where $d$ is the retriever's final hidden dimension. Importantly, we make no assumptions about how $R$ is normally scored at inference time: the process is the same whether $R$ is a single-vector retriever, where individual tokens would be pooled into one document-level vector, or a multi-vector model. In all cases, \method{} reads activations from the final hidden states of the backbone model.

We train an SAE on token-level activations drawn from $R$ run over unlabeled web text. Specifically, we sample passages from
FineWeb-Edu~\cite{finewebedu}. Every resulting token activation $h_i$ is treated as an independent training example for the SAE. Training minimizes the standard reconstruction objective discussed in Section~\ref{sec:background-sae}.

During training, we set the SAE's total latent vocabulary dimension to 32,768 terms, within the same order of magnitude as common monolingual tokenizers~\cite{bert}, and fix the top-k sparsity to 16 active features per token to ensure that the representations we will use for retrieval will naturally remain sparse. We found that increasing the latent vocabulary size, training data volume, or training-time top-k did not meaningfully improve downstream retrieval, which was robust to these hyperparameter choices overall. We provide further details on hyperparameters in Appendix~\ref{app:SAE}. Importantly, the SAE never sees data which is directly in-domain for retrieval tasks.

Following this training, the trained SAE encoder $f_{\mathrm{enc}}$ is frozen. It is used to project individual token into the 32,768 \emph{latent vocabulary} $V_{\mathrm{SAE}} = \{1, \ldots, m\}$ which will serve as our retrieval vocabulary in subsequent steps. The decoder $f_{\mathrm{dec}}$ plays no further role and is discarded.

\subsection{Constructing Latent Sparse Representations}
\label{sec:method-encoding}

At indexing and query time, we use $f_{\mathrm{enc}}$ as a token-level sparse projector. Given an input $x$, which can be either a document to be indexed or a query, we first obtain its token-level dense activations from $R$, then apply the SAE encoder independently at each position, relying on the backbone model's own contextualization of each token:
\begin{equation}
    z_i = f_{\mathrm{enc}}(h_i) \in \mathbb{R}^{m}_{\geq 0} .
\end{equation}
Each $z_i$ is sparse by construction, and the $m$ coordinates of $z_i$ are entries in the fixed \emph{latent vocabulary} $V_{\mathrm{SAE}} = \{1, \ldots, m\}$ learned by the SAE.

To produce a single representation per input, we aggregate the per-token codes by sum-pooling. Our experiments confirmed that max-pooling resulted in consistent minor performance degradation compared to sum-pooling. We believe this to be due to sum-pooling preserving the cumulative evidence contributed by repeated feature activations across the input, while max-pooling retains only the single strongest activation of each feature, discarding repeated weaker activations that contribute useful evidence. After pooling, we finally apply an element-wise activation transform $\phi: \mathbb{R}_{\geq 0} \to \mathbb{R}_{\geq 0}$:
\begin{equation}
    \tilde{w}(x)
    =
    \sum_{i=1}^{|x|}
    z_i,
    \qquad
    w(x)
    =
    \phi\!\left(
        \tilde{w}(x)
    \right)
    \in
    \mathbb{R}^{m}_{\geq 0} ,
\end{equation}

For $\phi$ we consider sublinear transforms such as $\phi(u) = u^{\alpha}$ for $\alpha \in (0, 1)$. This activation transform is beneficial because the summed SAE activations $\tilde{w}_j(d) = \sum_i z_{i,j}$ fundamentally differ from the term frequencies in natural language, where each $\operatorname{tf}_j(d)$ is a count. On the other hand, each per-token activation $z_{i,j}$ produced by the SAE encoder is a non-negative real-valued magnitude carrying signal rather than a simple indicator of token presence. For all experiments, we use the square-root transform $\phi(u)=\sqrt{u}$ as the default parameter, and otherwise include $\phi$ in our hyperparameter tuning process as described in Section~\ref{sec:tunedbm25}.

The resulting representation is a sparse, non-negative vector
$w(x) \in \mathbb{R}^{m}_{\geq 0}$ per input. Its support
\begin{equation}
    \operatorname{supp}(w(x)) = \{ j \in V_{\mathrm{SAE}} : w_j(x) > 0 \}
\end{equation}
identifies the features activated in $x$, while the magnitudes $w_j(x)$ capture the $\phi$-transformed activation strength of each feature across all tokens of $x$.

\subsection{BM25 Scoring over Latent Features}
\label{sec:method-scoring}

Finally, at retrieval time, given the sparse representations $w(q)$ and $w(d)$ defined above, \method{} scores query-document pairs by applying the Okapi BM25 formula over the latent vocabulary
$V_{\mathrm{SAE}}$. One adjustment to the lexical formulation is needed: since $w_j(q)$ is a real-valued activation rather than a binary indicator of term presence, we retain it as an explicit per-feature weight on each summand of the BM25 sum. Effectively, in the lexical case,
$j$ corresponds to a term and $w_j(D)$ is its term frequency in $D$; in our \method{} setting, we interpret the BM25 frequency term as $f(j,D) = w_j(D)$. We otherwise perform scoring and the inverse document frequency (IDF) calculation over the term activations as in Section~\ref{sec:background-bm25}.

In this context, all of BM25's structural mechanisms, saturating term-frequency contributions, document-length normalization, and IDF-based downweighting of pervasive features, transfer largely unchanged from natural language to \method{}. Thus, indexing and retrieval with this method is ``plug-and-play'' with existing infrastructure and can be carried out with any standard BM25 implementation that accepts a custom vocabulary: the latent feature indices serve directly as vocabulary entries in an inverted index.

\section{Experimental Setup}

\subsection{SAE Training}

\textbf{Training Setting} The SAEs are all trained using the same method mimicking standard best practices, following the Top-K SAE architecture introduced by~\citet{topksae}. We did not find any significant downstream improvements with SAE variants such as JumpReLU~\cite{jumprelu} or BatchTopK~\cite{batchtopk}. The decoder is initialized with Kaiming initialization and the encoder is initialized with the transposed weights of the decoder, following~\citet{SAEinterp}.  We use the AdamW optimizer with a maximum learning rate of $0.001$ with 5\% linear warmup followed by a cosine decay to 0. All trainings are performed on a single A100 GPU, taking under two hours per run. All SAEs are trained 5 times with different random seeds to minimise variance, with results reported as the average of these 5 runs.

\textbf{Dense Encoder Backbones.} We apply our method to three models to showcase its applicability to encoders with different training methods and scoring functions. Specifically, we use Contriever~\cite{contriever}\footnote{Specifically, the nthakur/contriever-base-msmarco available on the HuggingFace hub, as there appears to be multiple variants with conflicting reported results.}, a widely studied single-vector model, contemporary with SPLADE-v2~\cite{spladev2}. We also evaluate the single-vector retrieval model nomic-embed-text-v1.5~\cite{nomicembed} (\emph{\textbf{Nomic}}), a model using more modern training methods, contemporary to SPLADE-v3~\cite{spladev3} and with much stronger downstream performance than Contriever, to ensure that our method's gains are not restricted to weaker models. Finally, we also use GTE-ModernColBERT~\cite{gtemoderncolbert} (\emph{\textbf{GTE-MC}}), a strong multi-vector model following ColBERT~\cite{colbert}.\\

\subsection{Retrieval Evaluation Setup}

\paragraph{Baselines.} We report the results of various baselines to contextualise our method's performance. Our sparse baselines are lexical BM25 as well as multiple generations of SPLADE models: SPLADE-v2, SPLADE-v2-Distill, and SPLADE-v3. We also report the results of all three of our chosen backbone models evaluated in their normal scoring setting, and of \method{} over a non-finetuned BERT~\cite{bert} to confirm that our approach requires structures learned during retrieval training. 
\\
\textbf{Main Evaluation Data.} We report our main results over standard information retrieval benchmarks. Specifically, our main results are obtained by evaluating all methods on the widely used \textbf{BEIR}~\cite{beir} evaluation suite, containing 15 datasets across a variety of domains and which is currently the de facto standardised way of evaluating English information retrieval models.\\
\textbf{LIMIT.} To further understand the information extracted by our \method{} method, we also report results on LIMIT~\cite{limit}, a benchmark specifically designed to test the theoretical limitations of single-vector retrieval while being trivial for lexical models: while the strongest single-vector models score under 10\% on its main metric, Recall@20, BM25 reaches a score of above 95\%.

\subsubsection{BM25 Tuning} 
\label{sec:tunedbm25}

BM25 introduces two main tunable parameters, controlling term frequency penalties and length regularization, which are known to have a potentially substantial impact on retrieval performance~\cite{piagent,bm25optim}, and \method{} additionally introduces the tunable knob of the $\phi$ transform applied to raw model activations. For best results, it is common practice to tune BM25 to individual datasets to best match their idiosyncrasies~\cite{piagent,bm25optim2}. We find that \method{} is very resilient to BM25 hyperparameter choices, with Appendix~\ref{app:tuning} presenting a full comparison of results with and without tuning.

\section{Results}

\subsection{Main Retrieval Results}

\begin{table*}[t]
\centering
\scriptsize
\setlength{\tabcolsep}{1.55pt}
\renewcommand{\arraystretch}{1.08}
\resizebox{\textwidth}{!}{%
\begin{tabular}{@{}l*{16}{r}@{}}
\toprule
Method & Avg & SciF. & NFC & Arg. & SciD. & FiQA & TREC & Touch\'e & Quora & DBP & NQ & Clim. & Hotpot & FEVER & CQA & MSM \\
\midrule\midrule
BM25 & 0.419 & 0.686 & 0.319 & 0.399 & 0.163 & 0.249 & 0.680 & 0.667 & 0.804 & 0.300 & 0.285 & 0.136 & 0.569 & 0.481 & 0.314 & 0.234 \\
SPLADE-v2 & 0.450 & 0.628 & 0.313 & 0.439 & 0.145 & 0.287 & 0.673 & 0.316 & 0.835 & 0.366 & 0.469 & 0.199 & 0.636 & 0.730 & 0.311 & 0.403 \\
SPLADE-v2-distil & 0.490 & 0.693 & 0.334 & 0.479 & 0.158 & 0.336 & 0.710 & 0.364 & 0.838 & 0.435 & 0.521 & 0.235 & 0.684 & 0.786 & 0.350 & 0.434 \\
SPLADE-v3 & 0.502 & 0.710 & 0.357 & 0.509 & 0.158 & 0.374 & 0.748 & 0.293 & 0.814 & 0.450 & 0.586 & 0.233 & 0.692 & 0.796 & 0.344 & 0.469 \\
\midrule
Contriever & 0.415 & 0.655 & 0.313 & 0.484 & 0.171 & 0.274 & 0.448 & 0.153 & 0.867 & 0.377 & 0.419 & 0.248 & 0.542 & 0.581 & 0.334 & 0.359 \\
Nomic & 0.521 & 0.703 & 0.346 & 0.481 & 0.187 & 0.377 & 0.822 & 0.298 & 0.850 & 0.431 & 0.598 & 0.412 & 0.672 & 0.813 & 0.394 & 0.424 \\
GTE-MC & 0.547 & 0.756 & 0.381 & 0.477 & 0.192 & 0.456 & 0.849 & 0.316 & 0.867 & 0.475 & 0.617 & 0.306 & 0.773 & 0.875 & 0.410 & 0.453 \\
\midrule
BERT+\method{} & 0.245 & 0.585 & 0.216 & 0.206 & 0.103 & 0.131 & 0.212 & 0.151 & 0.711 & 0.134 & 0.165 & 0.116 & 0.345 & 0.283 & 0.240 & 0.083 \\
\midrule
Contriever+\method{} & \textit{0.474} & \textit{0.713} & \textit{0.340} & 0.436 & 0.165 & \textit{0.317} & \textit{0.709} & \textit{0.357} & 0.860 & \textit{0.409} & \textit{0.468} & \textit{0.254} & \textit{0.627} & \textit{0.751} & \textit{0.347} & \textit{0.366} \\
Nomic+\method{} & \textit{0.526} & \textit{0.749} & \textit{0.372} & 0.462 & 0.184 & \textit{0.382} & 0.783 & \textit{0.315} & \textit{0.855} & \textit{0.436} & 0.577 & 0.350 & \textit{0.732} & \textit{0.885} & \textit{0.394} & 0.417 \\
GTE-MC+\method{} & 0.500 & 0.730 & 0.374 & \textit{0.479} & 0.171 & 0.399 & 0.759 & \textit{0.405} & 0.865 & 0.387 & 0.509 & \textit{0.282} & 0.653 & 0.814 & 0.349 & 0.325 \\
\bottomrule
\end{tabular}%
}
\vspace{0.25em}
\caption{Main retrieval results. Italicized values in \method{} rows indicate that the \method{} variant improves over its base retriever. BM25-based methods reported with tuned parameters. All results reported as nDCG@10.}
\label{tab:mainresults}
\end{table*}

We present our main results on the full BEIR collection in Table~\ref{tab:mainresults}. Overall, we observe that \method{}, while fully out-of-domain, is a capable retriever across all evaluated settings. While it lags behind MaxSim scoring used by GTE-ModernColBERT, it outperforms the native cosine similarity scoring for both single-vector models evaluated. As expected, performance is particularly weak when used with an unfinetuned BERT, indicating that the necessary information is learned during retrieval fine-tuning.\\
\textbf{Comparison with SPLADE.} When paired with a backbone from the same era as a given SPLADE variant, \method{} consistently outperforms it. Combined with Nomic, it outperforms SPLADE-v3, while it outperforms the no-knowledge distillation variant of SPLADE-v2 when paired with Contriever.  Interesting dataset-level differences are noticeable: on domain-specific tasks such as FiQA or TREC-Covid, comparable \method{} results strongly outperform SPLADE variants. However, on ArguAna,  an argument mining task where lexical overlap is particularly important, SPLADE outperforms it, and the gap in performance is also significantly narrower on NQ, a large-scale QA dataset, again characterized by strong lexical overlap between questions and answers.\\
\textbf{Comparison with Dense Models.} The comparison of \method{} approaches with their dense backbones in their default scoring setting reveals some interesting patterns. As commonly thought, it appears that ColBERT's MaxSim scoring allows more of the model's learned knowledge to be expressed via its scoring mechanism, and thus remains noticeably stronger than its \method{} variant. On the other hand, \method{} is strong against both single-vector backbones, on both a weaker model such as Contriever or a competitive, near-state-of-the-art model like Nomic, but the magnitude of the performance differences is notable: while \method{}+Nomic only very slightly outperforms its backbone, essentially just matching its overall performance with different strengths and weaknesses, \method{}+Contriever is vastly superior to its native scoring setting. We hypothesize that Contriever's comparatively lighter training regimen produces good latent representations but does not fully saturate the model's final scoring pathway, leaving room for sparse extraction to recover additional signal. We hypothesize that Contriever's comparatively lighter training regimen produces good latent representations but does not fully exploit the model's final scoring pathway, leaving room for sparse extraction to recover additional signal. We intend to further explore this in future work.

\subsection{LIMIT}

\begin{table}[t]
\centering
\scriptsize
\begin{tabular}{@{}lrrrr@{}}
\toprule
Method & R@10 & R@20 & R@100 & R@1000 \\
\midrule
\midrule
BM25 & 0.9440 & 0.9490 & 0.9645 & 0.9945 \\
SPLADE-v3 & 0.5760 & 0.6650 & 0.8095 & 0.9440 \\
\midrule
GTE-MC & 0.8430 & 0.8565 & 0.8720 & 0.8795 \\
Contriever & 0.0210 & 0.0265 & 0.0530 & 0.1250 \\
\midrule
GTE-MC+\method{} & 0.7985 & 0.8315 & 0.8915 & 0.9775 \\
Contriever+\method{} & 0.4140 & 0.5100 & 0.7295 & 0.9290 \\
\bottomrule
\end{tabular}
\vspace{0.25em}
\caption{Recall@k of selected models on LIMIT.}
\label{tab:limit}
\end{table}

Next, Table~\ref{tab:limit} presents the results of selected methods on LIMIT. On common retrieval methods, our results reproduce those of the original paper: BM25 reaches extremely strong performance, closely followed by multi-vector retrieval models, with SPLADE models reaching weaker results and single-vector models completely collapsing on this task. This is line with what the paper introducing LIMIT proposes: a deliberately simple task with noisy lexical attributes designed to crowd out signal in single-vector representations.

\method{} appears to significantly recover performance, reaching a score over 25 times higher when applied to Contriever compared to its single-vector setting. This does offer strong insight further supporting the idea that the inherent limits of single-vector retrievers lie in their scoring mechanism, \emph{but} that this scoring mechanism also offers sufficient training signal for the underlying model to learn to generate representations that are able to at least partially capture this signal. With the GTE-MC backbone, MaxSim once again allows it to be the strongest neural model evaluated by directly leveraging token-level signal. However, it appears that while its dense representation mechanism has better ranking abilities than \method{} applied over it, it also reaches earlier recall saturation: its Recall@1000 caps out at 87.95\%, almost identical to its Recall@100, whereas 
\method{} reaches 97.75\%, suggesting better long-tail performance and virtually matching purely lexical approaches.

\subsection{Overall}

\begin{figure}
    \centering
    \includegraphics[width=0.9\linewidth]{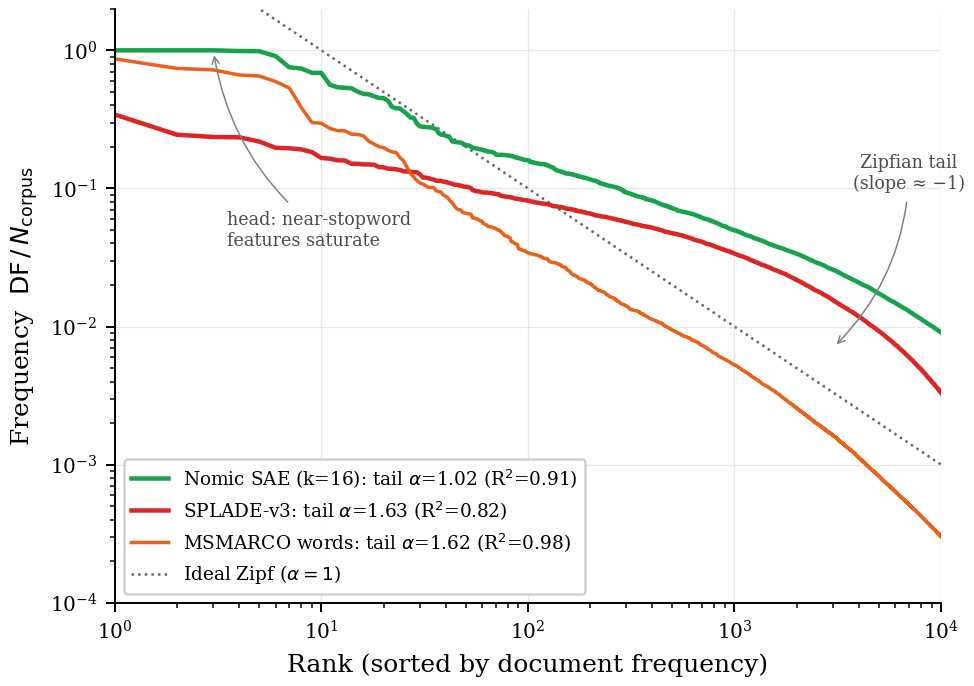}
    \caption{Frequency distribution of activated features.}
    \label{fig:zipf}
\end{figure}

Overall, these results appear to suggest that \method{} is able to extract a retrieval-native vocabulary from dense retrievers, and that this vocabulary can be used as a capable retriever when combined with traditional IR methods for handling sparse representations.
We believe that the overall strength of MaxSim on all tasks, and the results of \method{} variants on both classical tasks such as BEIR, and LIMIT, where it recovers performance that is otherwise collapsed in single-vector scoring methods both reinforce our hypothesis: Dense retrievers learn expressive representations, but there exist cases where they cannot be expressed in a way that is captured by their scoring mechanism.

\section{The Anatomy of \method{}}

\begin{table*}[t]
\centering
\small
\setlength{\tabcolsep}{6pt}
\renewcommand{\arraystretch}{1.25}

\newcommand{\fid}[1]{\texttt{\textbf{#1}}}
\newcommand{\act}[1]{\textbf{#1}}

\begin{tabularx}{\linewidth}{@{}p{0.08\linewidth} p{0.25\linewidth} X@{}}
\toprule
\textbf{Type}
&
\textbf{Captured Information}
&
\textbf{Representative activations}
\\
\midrule

\textbf{Lexical}
&
A surface-form feature, regardless of surrounding context.
&
\fid{F4384} \emph{bridge}: 
Covered \act{Bridge}; Moore's Creek \act{Bridge}; Mount Hope \act{Bridge}.

\fid{F2196} \emph{normal}: 
\act{normal}, healthy person; \act{Normal} response; \act{normal} physiology.
\\
\midrule

\textbf{Narrow Semantic}
&
A concept-level feature: close synonyms, inflections, or paraphrases of the same idea.
&
\fid{F17567} \emph{buy / purchase}: 
\act{buy} a Treasury bill; can be \act{bought}; satisfied with your \act{purchase}.

\fid{F305} \emph{football roles}: 
Welsh \act{footballer}; plays as a \act{striker}; highest-paid \act{player}.
\\
\midrule

\textbf{Broad Topical}
&
A topic-level feature: lexically different words that belong to the same broader domain.
&
\fid{F21138} \emph{ancient / archaeology}: 
\act{prehistoric} times; \act{archaeological} park; \act{ancient} Egyptians.

\fid{F19067} \emph{software dev.}: 
\act{software} engineer; junior \act{developer}; source \act{code}.
\\

\bottomrule
\end{tabularx}

\caption{
Representative features sampled from each of the three qualitatively identified categories.}
\label{tab:feature-types}
\end{table*}


\subsection{SAE Features Have Term-Like Collection Statistics}
\label{sec:zipf}

We now explore why BM25 works out-of-the-box on the generated sparse features while dot product scoring does not, in complete contrast with SPLADE models, for which BM25 scoring yields significantly degraded results (Appendix~\ref{app:bm25splade}).

We attribute this to the distributional properties of SAE features. Indeed, BM25 was developed specifically to match the distributional properties of natural language, which is understood to follow a quasi-Zipfian distribution~\cite{worddistrib}, meaning that the $r-th$ most common word appears roughly $1/r$ times as often as the most common one~\cite{zipf}.
This shape is key to enabling the weighting parameters of BM25 to act as an effective way to discriminate between documents.

Figure~\ref{fig:zipf} presents the term distribution over the full MS MARCO~\cite{msmarco} collection for SPLADE-v3, \method{}+Nomic and natural language. The features generated by SPLADE diverge from Zipf's law via a lack of dominant, quasi-stopword features, common in natural language. On the other hand, SAE-generated features, while not a perfect fit, are Zipfian in nature. Notably, they generated very pronounced saturated terms at the top of the distribution before adopting a curve that remains less steep than that of natural terms', which is seemingly sufficient to leverage BM25's weighting mechanisms. Appendix~\ref{app:pruning} presents a rapid exploration of saturated term pruning.

\begin{figure}
    \centering
    \includegraphics[width=0.95\linewidth]{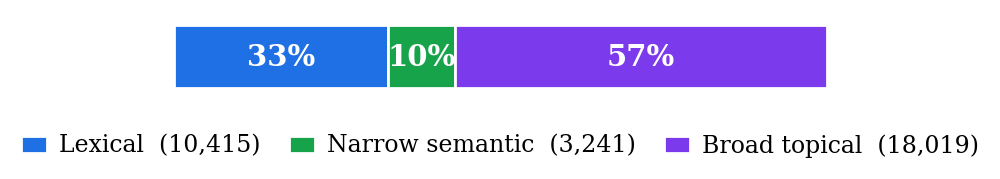}
    \caption{Distribution of features by feature types.}
    \label{fig:distrib}
\end{figure}

\subsection{What Do Latent Retrieval Terms Capture?}

Next, we qualitatively identify three categories of features, presented with examples in Table~\ref{tab:feature-types}, and design a simple automated annotation process during which we use Gemini 3 Pro~\cite{gemini3} to annotate all vocabulary terms. We then randomly sample 500 of its annotations and manually review them, finding perfect human-LLM agreement.
The results of this annotation process are presented in  Figure~\ref{fig:distrib}. The majority of features fall within the \textbf{broad topical} category, with a third of the features being lexical and just 10\% being narrow semantic ones. This distribution suggests that the information captured by \method{} tend towards a form of "hybrid" semantic-lexical representation, with around two-thirds of its features being primarily semantic and the remaining third focusing on purely lexical matches. Interestingly, this falls in line with the existing literature, which has long argued that purely semantic matching misses information~\cite{dpr} that can be recovered through hybrids of semantic and lexical methods~\cite{rrf}.

\section{Conclusion}

In this paper, we introduce \textbf{\method{}}, which demonstrates that dense retrievers contain more information than just the than what is exposed through their cosine similarity-based scoring mechanism. These features can be extracted by Sparse Autoencoders without any retrieval-specific modifications, yielding extracted features that are Zipfian in nature and approach the distribution of natural language. We further show that these features are suitable for BM25 scoring, originally designed for lexical terms, and can reach retrieval performance that surpasses that of the same model used in its native single-vector similarity setting. Finally, qualitative analysis reveals that these features capture multiple categories of information, creating a ``hybrid'' mix of lexical and semantic features. These results are obtained without any retrieval-specific data during the training of the SAE, highlighting that dense retrievers naturally learn meaningful, indexable sparse representations. We believe that these results encourage future research into decoupling the study of scoring operators from that of retrieval representation learning to better understand what truly limits the expressivity of retrievers.

\section*{Limitations}

We identify six main limitations to our study, which we plan to address in future work.

\textbf{Language. } Our study largely focuses on the English language, as it is the highest resource language to demonstrate the mechanisms studied. We believe future work should extend this approach to both non-English monolingual and multilingual models, as it is possible that a shared, cross-lingual vocabulary could be surfaced by the SAE encoder.

\textbf{Lexical/Semantic Hybridification. } While our analysis reveals that the features hybridize to an extent, our results do not appear to fully match the strength of a true Dense + Sparse hybrid method as it suffers from some drawbacks on datasets where one or the other is typically strong. However, we believe that our results are encouraging and point towards \method{} potentially paving the way to better hybridification of feature types which warrants further exploration.

\textbf{SAE Variants and SAE Limitations. } We have two limitations related to SAEs: The first is that while we did not find BatchTopK~\cite{batchtopk} or JumpReLU~\cite{jumprelu} to outperform our Top-K SAE~\cite{topksae}, the SAE literature is growing rapidly. Whether some other SAE variant is better suited to extracting retrieval-adapted features remains an open question. Additionally, while SAEs are popular models, they are known to have inherent limitations that have not yet been overcome~\cite{saelimits}, notably around feature completeness~\cite{leanske} and potential dependency on the training dataset~\cite{kissane}.

\textbf{Sparsity. } This study does not deeply explore how the sparsity generated by our method actually manifests, and whether there are techniques to make it more efficient, such as by eliminating saturated terms from the index as proposed by~\citet{spladeefficient} to increase the efficiency of SPLADE, especially as Figure~\ref{fig:zipf} appears to indicate there exists many such terms.

\textbf{Alternate Scoring Approaches. } Our approach demonstrates that dense models contain extractable sparse features, using BM25 as the scoring mechanism. However, BM25 is just one of many scoring methods, and although it is empirically strong for lexical terms, future work should explore scoring methods that could be better suited to \method{}.

\textbf{\method{}+ColBERT as a separate class. } In this study, we explore the use of our method applied to one ColBERT model, but otherwise treat the late interaction family of retrieval models as a special class of dense retrievers. While this is taxonomically reasonable, ColBERT's strong token-level signal may warrant dedicated extraction methods that exploit late-interaction structure more directly.

\section*{Ethical Considerations}

All retrieval models are currently understood to contain poorly-understood biases, and can potentially result in downstream issues should they surface such biased results which are not understood be biased. While we believe this to be an issue that warrants further work to be alleviated, our work, focusing on extracting representations within existing models, does not meaningfully carry considerably greater risk than existing retrieval methods. Due to its non-generative nature, we believe that our work is unlikely to be able to result in significant harm.

\bibliography{custom}

@inproceedings{bert,
  title={BERT: Pre-training of Deep Bidirectional Transformers for Language Understanding},
  author={Devlin, Jacob and Chang, Ming-Wei and Lee, Kenton and Toutanova, Kristina},
  booktitle={Proceedings of the 2019 Conference of the North American Chapter of the Association for Computational Linguistics: Human Language Technologies, Volume 1 (Long and Short Papers)},
  pages={4171--4186},
  year={2019}
}

@article{bm25,
  title={Okapi at TREC-3},
  author={Robertson, Stephen E and Walker, Steve and Jones, Susan and Hancock-Beaulieu, Micheline M and Gatford, Mike and others},
  journal={Nist Special Publication Sp},
  volume={109},
  pages={109},
  year={1995},
  publisher={National Instiute of Standards \& Technology}
}

@article{beir,
  title={Beir: A heterogenous benchmark for zero-shot evaluation of information retrieval models},
  author={Thakur, Nandan and Reimers, Nils and R{\"u}ckl{\'e}, Andreas and Srivastava, Abhishek and Gurevych, Iryna},
  journal={arXiv preprint arXiv:2104.08663},
  year={2021}
}

@inproceedings{splade,
  title={SPLADE: Sparse lexical and expansion model for first stage ranking},
  author={Formal, Thibault and Piwowarski, Benjamin and Clinchant, St{\'e}phane},
  booktitle={Proceedings of the 44th International ACM SIGIR Conference on Research and Development in Information Retrieval},
  pages={2288--2292},
  year={2021}
}

@String{Computing = "Computing" }

@String{Springer = "Springer-Verlag" }

@inproceedings{colbert,
  title={Colbert: Efficient and effective passage search via contextualized late interaction over bert},
  author={Khattab, Omar and Zaharia, Matei},
  booktitle={Proceedings of the 43rd International ACM SIGIR conference on research and development in Information Retrieval},
  pages={39--48},
  year={2020}
}

@article{spladev3,
  title={SPLADE-v3: New baselines for SPLADE},
  author={Lassance, Carlos and D{\'e}jean, Herv{\'e} and Formal, Thibault and Clinchant, St{\'e}phane},
  journal={arXiv preprint arXiv:2403.06789},
  year={2024}
}

@article{msmarco,
  title={{MS MARCO}: A human generated machine reading comprehension dataset},
  author={Nguyen, Tri and Rosenberg, Mir and Song, Xia and Gao, Jianfeng and Tiwary, Saurabh and Majumder, Rangan and Deng, Li},
  journal={choice},
  volume={2640},
  pages={660},
  year={2016}
}

@inproceedings{dpr,
  title={Pretrained transformers for text ranking: BERT and beyond},
  author={Yates, Andrew and Nogueira, Rodrigo and Lin, Jimmy},
  booktitle={Proceedings of the 14th ACM International Conference on web search and data mining},
  pages={1154--1156},
  year={2021}
}

@inproceedings{leanske,
  title={Sparse autoencoders do not find canonical units of analysis},
  author={Leask, Patrick and Bussmann, Bart and Pearce, Michael and Bloom, Joseph and Tigges, Curt and Al Moubayed, Noura and Sharkey, Lee and Nanda, Neel},
  booktitle={International Conference on Learning Representations},
  volume={2025},
  pages={53617--53642},
  year={2025}
}

@inproceedings{kissane,
  title={Saes are highly dataset dependent: A case study on the refusal direction},
  author={Kissane, Connor and Krzyzanowski, Robert and Nanda, Neel and Conmy, Arthur},
  booktitle={Alignment Forum},
  year={2024}
}

@inproceedings{spladeefficient,
author = {Lassance, Carlos and Clinchant, St\'{e}phane},
title = {An Efficiency Study for SPLADE Models},
year = {2022},
isbn = {9781450387323},
publisher = {Association for Computing Machinery},
address = {New York, NY, USA},
url = {https://doi.org/10.1145/3477495.3531833},
doi = {10.1145/3477495.3531833},
booktitle = {Proceedings of the 45th International ACM SIGIR Conference on Research and Development in Information Retrieval},
pages = {2220–2226},
numpages = {7},
location = {Madrid, Spain},
series = {SIGIR '22}
}

@article{spladev2,
  title={{SPLADE} v2: Sparse lexical and expansion model for information retrieval},
  author={Formal, Thibault and Lassance, Carlos and Piwowarski, Benjamin and Clinchant, St{\'e}phane},
  journal={arXiv preprint arXiv:2109.10086},
  year={2021}
}

@article{contriever,
  title={Unsupervised Dense Information Retrieval with Contrastive Learning},
  author={Izacard, Gautier and Caron, Mathilde and Hosseini, Lucas and Riedel, Sebastian and Bojanowski, Piotr and Joulin, Armand and Grave, Edouard},
  journal={Transactions on Machine Learning Research},
  year={2022},
}

@misc{jacolbertv2.5,
      title={JaColBERTv2.5: Optimising Multi-Vector Retrievers to Create State-of-the-Art Japanese Retrievers with Constrained Resources}, 
      author={Benjamin Clavié},
      year={2024},
      eprint={2407.20750},
      archivePrefix={arXiv},
      primaryClass={cs.IR},
      url={https://arxiv.org/abs/2407.20750}, 
}

@article{neuralir,
author = {Miutra, Bhaskar and Craswell, Nick},
title = {An Introduction to Neural Information Retrieval },
year = {2018},
issue_date = {Dec 2018},
publisher = {Now Publishers Inc.},
address = {Hanover, MA, USA},
volume = {13},
number = {1},
issn = {1554-0669},
url = {https://doi.org/10.1561/1500000061},
doi = {10.1561/1500000061},
journal = {Found. Trends Inf. Retr.},
month = dec,
pages = {1–126},
numpages = {129}
}

@misc{gtemoderncolbert,
title={{GTE-ModernColBERT}},
author={Chaffin, Antoine},
url={https://huggingface.co/lightonai/GTE-ModernColBERT-v1},
year={2025},
note={HuggingFace Hub}
}

@inproceedings{flopsalt,
author = {Porco, Aldo and Mehra, Dhruv and Malioutov, Igor and Radhakrishnan, Karthik and Keymanesh, Moniba and Preo\c{t}iuc-Pietro, Daniel and MacAvaney, Sean and Cheng, Pengxiang},
title = {An Alternative to FLOPS Regularization to Effectively Productionize SPLADE-Doc},
year = {2025},
isbn = {9798400715921},
publisher = {Association for Computing Machinery},
address = {New York, NY, USA},
url = {https://doi.org/10.1145/3726302.3730163},
doi = {10.1145/3726302.3730163},
booktitle = {Proceedings of the 48th International ACM SIGIR Conference on Research and Development in Information Retrieval},
pages = {2789–2793},
numpages = {5},
location = {Padua, Italy},
series = {SIGIR '25}
}

@misc{gemmascope,
      title={Gemma Scope: Open Sparse Autoencoders Everywhere All At Once on Gemma 2}, 
      author={Tom Lieberum and Senthooran Rajamanoharan and Arthur Conmy and Lewis Smith and Nicolas Sonnerat and Vikrant Varma and János Kramár and Anca Dragan and Rohin Shah and Neel Nanda},
      year={2024},
      eprint={2408.05147},
      archivePrefix={arXiv},
      primaryClass={cs.LG},
      url={https://arxiv.org/abs/2408.05147}, 
}

@article{anthro,
  title={Scaling monosemanticity: Extracting interpretable features from {C}laude 3 {S}onnet},
  author={Templeton, Adly and Conerly, Tom and Marcus, Jonathan and Lindsey, Jack and Bricken, Trenton and Chen, Brian and Pearce, Adam and Citro, Craig and Ameisen, Emmanuel and Jones, Andy and others},
  journal={Transformers Circuits},
  year={2025}
}

@article{SAEinterp,
  title={Towards monosemanticity: Decomposing language models with dictionary learning},
  author={Bricken, Trenton and Templeton, Adly and Batson, Joshua and Chen, Brian and Jermyn, Adam and Conerly, Tom and Turner, Nick and Anil, Cem and Denison, Carson and Askell, Amanda and others},
  journal={Transformer Circuits},
  year={2023}
}

@inproceedings{deepimpact,
author = {Mallia, Antonio and Khattab, Omar and Suel, Torsten and Tonellotto, Nicola},
title = {Learning Passage Impacts for Inverted Indexes},
year = {2021},
isbn = {9781450380379},
publisher = {Association for Computing Machinery},
address = {New York, NY, USA},
url = {https://doi.org/10.1145/3404835.3463030},
doi = {10.1145/3404835.3463030},
booktitle = {Proceedings of the 44th International ACM SIGIR Conference on Research and Development in Information Retrieval},
pages = {1723–1727},
numpages = {5},
location = {Virtual Event, Canada},
series = {SIGIR '21}
}

@inproceedings{clsr,
    title = "Decoding Dense Embeddings: Sparse Autoencoders for Interpreting and Discretizing Dense Retrieval",
    author = "Park, Seongwan  and
      Kim, Taeklim  and
      Ko, Youngjoong",
    editor = "Christodoulopoulos, Christos  and
      Chakraborty, Tanmoy  and
      Rose, Carolyn  and
      Peng, Violet",
    booktitle = "Proceedings of the 2025 Conference on Empirical Methods in Natural Language Processing",
    month = nov,
    year = "2025",
    address = "Suzhou, China",
    publisher = "Association for Computational Linguistics",
    url = "https://aclanthology.org/2025.emnlp-main.1345/",
    doi = "10.18653/v1/2025.emnlp-main.1345",
    pages = "26468--26485",
    ISBN = "979-8-89176-332-6"
}

@inproceedings{finewebedu,
author = {Penedo, Guilherme and Kydl\'{\i}\v{c}ek, Hynek and Allal, Loubna Ben and Lozhkov, Anton and Mitchell, Margaret and Raffel, Colin and Von Werra, Leandro and Wolf, Thomas},
title = {The FineWeb datasets: decanting the web for the finest text data at scale},
year = {2024},
isbn = {9798331314385},
publisher = {Curran Associates Inc.},
address = {Red Hook, NY, USA},
booktitle = {Proceedings of the 38th International Conference on Neural Information Processing Systems},
articleno = {970},
numpages = {39},
location = {Vancouver, BC, Canada},
series = {NIPS '24}
}

@inproceedings{doc2query,
author = {Gospodinov, Mitko and MacAvaney, Sean and Macdonald, Craig},
title = {Doc2Query–: When Less is More},
year = {2023},
isbn = {978-3-031-28237-9},
publisher = {Springer-Verlag},
address = {Berlin, Heidelberg},
url = {https://doi.org/10.1007/978-3-031-28238-6_31},
doi = {10.1007/978-3-031-28238-6_31},
booktitle = {Advances in Information Retrieval: 45th European Conference on Information Retrieval, ECIR 2023, Dublin, Ireland, April 2–6, 2023, Proceedings, Part II},
pages = {414–422},
numpages = {9},
location = {Dublin, Ireland}
}

@misc{topksae,
      title={Scaling and evaluating sparse autoencoders}, 
      author={Leo Gao and Tom Dupré la Tour and Henk Tillman and Gabriel Goh and Rajan Troll and Alec Radford and Ilya Sutskever and Jan Leike and Jeffrey Wu},
      year={2024},
      eprint={2406.04093},
      archivePrefix={arXiv},
      primaryClass={cs.LG},
      url={https://arxiv.org/abs/2406.04093}, 
}

@inproceedings{batchtopk,
title={BatchTopK Sparse Autoencoders},
author={Bart Bussmann and Patrick Leask and Neel Nanda},
booktitle={NeurIPS 2024 Workshop on Scientific Methods for Understanding Deep Learning},
year={2024},
url={https://openreview.net/forum?id=d4dpOCqybL}
}

@misc{faire,
  author = {Hu, Xiaomeng},
  title = {Beyond BM25 and Dense Embeddings: How We Built Smart and Interpretable Retrieval at Faire},
  year = {2026},
  month = mar,
  day = {31},
  url = {https://craft.faire.com/beyond-bm25-and-dense-embeddings-841a7b18ce27},
  note = {The Craft, Medium Post}
}

@article{jumprelu,
  title={Jumping ahead: Improving reconstruction fidelity with jumprelu sparse autoencoders},
  author={Rajamanoharan, Senthooran and Lieberum, Tom and Sonnerat, Nicolas and Conmy, Arthur and Varma, Vikrant and Kram{\'a}r, J{\'a}nos and Nanda, Neel},
  journal={arXiv preprint arXiv:2407.14435},
  year={2024}
}

@article{worddistrib,
  title={Zipf's law in 50 languages: its structural pattern, linguistic interpretation, and cognitive motivation},
  author={Yu, Shuiyuan and Xu, Chunshan and Liu, Haitao},
  journal={arXiv preprint arXiv:1807.01855},
  year={2018}
}

@article{zipf,
  title={Selected studies of the principle of relative frequency in language},
  author={Zipf, George Kingsley},
  year={1932},
  publisher={Harvard university press}
}

@inproceedings{rrf,
author = {Cormack, Gordon V. and Clarke, Charles L A and Buettcher, Stefan},
title = {Reciprocal rank fusion outperforms condorcet and individual rank learning methods},
year = {2009},
isbn = {9781605584836},
publisher = {Association for Computing Machinery},
address = {New York, NY, USA},
url = {https://doi.org/10.1145/1571941.1572114},
doi = {10.1145/1571941.1572114},
booktitle = {Proceedings of the 32nd International ACM SIGIR Conference on Research and Development in Information Retrieval},
pages = {758–759},
numpages = {2},
location = {Boston, MA, USA},
series = {SIGIR '09}
}

@techreport{gemini3,
  title       = {Gemini 3 Pro Model Card},
  author      = {{Google DeepMind}},
  institution = {Google DeepMind},
  type        = {Model card},
  year        = {2025},
  month       = dec,
  url         = {https://storage.googleapis.com/deepmind-media/Model-Cards/Gemini-3-Pro-Model-Card.pdf},
  note        = {Model release: November 2025. Accessed: 2026-05-22}
}

@inproceedings{unicoil,
    title = "{COIL}: Revisit Exact Lexical Match in Information Retrieval with Contextualized Inverted List",
    author = "Gao, Luyu  and
      Dai, Zhuyun  and
      Callan, Jamie",
    editor = "Toutanova, Kristina  and
      Rumshisky, Anna  and
      Zettlemoyer, Luke  and
      Hakkani-Tur, Dilek  and
      Beltagy, Iz  and
      Bethard, Steven  and
      Cotterell, Ryan  and
      Chakraborty, Tanmoy  and
      Zhou, Yichao",
    booktitle = "Proceedings of the 2021 Conference of the North American Chapter of the Association for Computational Linguistics: Human Language Technologies",
    month = jun,
    year = "2021",
    address = "Online",
    publisher = "Association for Computational Linguistics",
    url = "https://aclanthology.org/2021.naacl-main.241/",
    doi = "10.18653/v1/2021.naacl-main.241",
    pages = "3030--3042"
}

@article{spladepp,
author = {Formal, Thibault and Lassance, Carlos and Piwowarski, Benjamin and Clinchant, St\'{e}phane},
title = {Towards Effective and Efficient Sparse Neural Information Retrieval},
year = {2024},
issue_date = {September 2024},
publisher = {Association for Computing Machinery},
address = {New York, NY, USA},
volume = {42},
number = {5},
issn = {1046-8188},
url = {https://doi.org/10.1145/3634912},
doi = {10.1145/3634912},
journal = {ACM Trans. Inf. Syst.},
month = apr,
articleno = {116},
numpages = {46}
}

@article{nomicembed,
title={Nomic Embed: Training a Reproducible Long Context Text Embedder},
author={Zach Nussbaum and John Xavier Morris and Andriy Mulyar and Brandon Duderstadt},
journal={Transactions on Machine Learning Research},
issn={2835-8856},
year={2025},
url={https://openreview.net/forum?id=IPmzyQSiQE},
note={Reproducibility Certification}
}

@inproceedings{limit,
title={On the Theoretical Limitations of Embedding-Based Retrieval},
author={Orion Weller and Michael Boratko and Iftekhar Naim and Jinhyuk Lee},
booktitle={The Fourteenth International Conference on Learning Representations},
year={2026},
url={https://openreview.net/forum?id=k9CzIvzfaA}
}

@misc{piagent,
      title={Rethinking Agentic Search with Pi-Serini: Is Lexical Retrieval Sufficient?}, 
      author={Tz-Huan Hsu and Jheng-Hong Yang and Jimmy Lin},
      year={2026},
      eprint={2605.10848},
      archivePrefix={arXiv},
      primaryClass={cs.IR},
      url={https://arxiv.org/abs/2605.10848}, 
}

@inproceedings{
splare,
title={Learning Retrieval Models with Sparse Autoencoders},
author={Thibault Formal and Maxime Louis and Herv{\'e} D{\'e}jean and St{\'e}phane Clinchant},
booktitle={The Fourteenth International Conference on Learning Representations},
year={2026},
url={https://openreview.net/forum?id=TuFjICawSc}
}

@article{deepct,
  title={Context-aware sentence/passage term importance estimation for first stage retrieval},
  author={Dai, Zhuyun and Callan, Jamie},
  journal={arXiv preprint arXiv:1910.10687},
  year={2019}
}

@article{doct5query,
  title={Document expansion by query prediction},
  author={Nogueira, Rodrigo and Yang, Wei and Lin, Jimmy and Cho, Kyunghyun},
  journal={arXiv preprint arXiv:1904.08375},
  year={2019}
}

@inproceedings{clip,
  title={Learning transferable visual models from natural language supervision},
  author={Radford, Alec and Kim, Jong Wook and Hallacy, Chris and Ramesh, Aditya and Goh, Gabriel and Agarwal, Sandhini and Sastry, Girish and Askell, Amanda and Mishkin, Pamela and Clark, Jack and others},
  booktitle={International conference on machine learning},
  pages={8748--8763},
  year={2021},
  organization={PmLR}
}

@misc{bm25v,
      title={Visual Words Meet BM25: Sparse Auto-Encoder Visual Word Scoring for Image Retrieval}, 
      author={Donghoon Han and Eunhwan Park and Seunghyeon Seo},
      year={2026},
      eprint={2603.05781},
      archivePrefix={arXiv},
      primaryClass={cs.CV},
      url={https://arxiv.org/abs/2603.05781}, 
}

@misc{SAE1,
      title={Sparse Autoencoders Find Highly Interpretable Features in Language Models}, 
      author={Hoagy Cunningham and Aidan Ewart and Logan Riggs and Robert Huben and Lee Sharkey},
      year={2023},
      eprint={2309.08600},
      archivePrefix={arXiv},
      primaryClass={cs.LG},
      url={https://arxiv.org/abs/2309.08600}, 
}

@article{saelimits,
title={Open Problems in Mechanistic Interpretability},
author={Lee Sharkey and Bilal Chughtai and Joshua Batson and Jack Lindsey and Jeffrey Wu and Lucius Bushnaq and Nicholas Goldowsky-Dill and Stefan Heimersheim and Alejandro Ortega and Joseph Isaac Bloom and Stella Biderman and Adri{\`a} Garriga-Alonso and Arthur Conmy and Neel Nanda and Jessica Mary Rumbelow and Martin Wattenberg and Nandi Schoots and Joseph Miller and William Saunders and Eric J Michaud and Stephen Casper and Max Tegmark and David Bau and Eric Todd and Atticus Geiger and Mor Geva and Jesse Hoogland and Daniel Murfet and Thomas McGrath},
journal={Transactions on Machine Learning Research},
issn={2835-8856},
year={2025},
url={https://openreview.net/forum?id=91H76m9Z94},
note={Survey Certification}
}

@inproceedings{
hardnegs,
title={Approximate Nearest Neighbor Negative Contrastive Learning for Dense Text Retrieval},
author={Lee Xiong and Chenyan Xiong and Ye Li and Kwok-Fung Tang and Jialin Liu and Paul N. Bennett and Junaid Ahmed and Arnold Overwijk},
booktitle={International Conference on Learning Representations},
year={2021},
url={https://openreview.net/forum?id=zeFrfgyZln}
}

@inproceedings{bm25optim,
  title={Which BM25 Do You Mean? A Large-Scale Reproducibility Study of Scoring Variants},
  author={Kamphuis, Chris and de Vries, Arjen P. and Boytsov, Leonid and Lin, Jimmy},
  booktitle={European Conference on Information Retrieval},
  pages={28--34},
  year={2020},
  doi={10.1007/978-3-030-45442-5_4}
}

@InProceedings{bm25optim2,
author="He, Ben
and Ounis, Iadh",
editor="Losada, David E.
and Fern{\'a}ndez-Luna, Juan M.",
title="Term Frequency Normalisation Tuning for BM25 and DFR Models",
booktitle="Advances in Information Retrieval",
year="2005",
publisher="Springer Berlin Heidelberg",
address="Berlin, Heidelberg",
pages="200--214",
isbn="978-3-540-31865-1"
}

\appendix

\section{SAE Parameters}
\label{app:SAE}

\begin{table*}[!ht]
\centering
\small
\setlength{\tabcolsep}{8pt}
\renewcommand{\arraystretch}{1.15}
\begin{tabular}{@{}ll@{}}
\toprule
\textbf{Parameter} & \textbf{Value} \\
\midrule
Architecture        & Top-K SAE~\cite{topksae} \\
Latent dim ($m$)    & 32{,}768 \\
activated $k$ (train)         & 16 \\
activated $k$ (inference)     & 16 \\
Encoder init        & Transposed decoder weights~\cite{SAEinterp} \\
Decoder init        & Kaiming \\
\midrule
Optimizer           & AdamW \\
Peak learning rate  & $10^{-3}$ \\
Batch Size          & 4{,}096 \\
Schedule            & Cosine decay to $0$ \\
Warmup              & 5\% linear \\
\midrule
Total unique tokens & 30B \\
Total epochs        & 3 \\
Hardware            & 1 $\times$ A100 \\
Corpus              & FineWeb-Edu (random sample) \\
\bottomrule
\end{tabular}
\caption{SAE training hyperparameters used for all \method{} runs reported in the main results.}
\label{tab:app-sae-params}
\end{table*}

The full parameters we used for the final model are presented in Table~\ref{tab:app-sae-params}.



\section{Term Pruning}
\label{app:pruning}

\begin{table*}[!ht]
\centering
\small
\setlength{\tabcolsep}{10pt}
\renewcommand{\arraystretch}{1.15}
\begin{tabular}{@{}lrrr@{}}
\toprule
\textbf{Prune \%} & \textbf{\# Pruned Feats.} & \textbf{nDCG@10} & \textbf{MRR@10} \\
\midrule
0\%  & 0       & 0.4095            & 0.3586 \\
1\%  & 327     & 0.4053 ($-1.0\%$) & 0.3546 ($-1.1\%$) \\
5\%  & 1{,}638 & 0.3904 ($-4.7\%$) & 0.3409 ($-4.9\%$) \\
10\% & 3{,}276 & 0.3755 ($-8.3\%$) & 0.3272 ($-8.8\%$) \\
\bottomrule
\end{tabular}
\caption{Effect of pruning the most-activated latent features on retrieval quality. Percentage changes relative to no pruning shown in parentheses.}
\label{tab:app-pruning}
\end{table*}

Section~\ref{sec:zipf} showed that \method{} features have a heavy head, with many features appearing to be saturated. This prompts investigation into whether or not pruning such terms would hurt. In Table~\ref{tab:app-pruning}, we present the performance impact of pruning the most frequent terms from the vocabulary prior to indexing on performance, on MS MARCO~\cite{msmarco}, the dataset used in Section~\ref{sec:zipf}.

We notice that performance appears to be robust to pruning the top 1\% of features, but that it otherwise rapidly degrades with more aggressive pruning. This suggests that while the most saturated terms appear to have little discriminative values, the rest of the heavy head of the distribution does play a discriminative role.
 
\section{Dot Product vs BM25 Scoring}
\label{app:bm25splade}

\begin{table}[!ht]
\centering
\scriptsize
\setlength{\tabcolsep}{4.5pt}
\renewcommand{\arraystretch}{1.08}
\begin{tabular}{@{}lrrrrrr@{}}
\toprule
Method & Avg & FiQA & TREC & NQ & Hotpot & DBP \\
\midrule
SPLADE-v3 dot & 0.568 & 0.380 & 0.732 & \textbf{0.587} & 0.692 & \textbf{0.450} \\
SPLADE-v3 BM25 & 0.455 & 0.318 & 0.583 & 0.417 & 0.594 & 0.362 \\
\method{} BM25 & \textbf{0.582} & \textbf{0.382} & \textbf{0.783} & 0.577 & \textbf{0.732} & 0.436 \\
\method{} dot & 0.381 & 0.194 & 0.747 & 0.273 & 0.372 & 0.321 \\
\bottomrule
\end{tabular}
\vspace{0.25em}
\caption{Comparison of dot-product and BM25 scoring on select datasets. All values report nDCG@10. Best values per dataset are bolded. \method{} uses Nomic in all cases.}
\label{tab:app-scoring}
\end{table}

In Table~\ref{tab:app-scoring}, we present the results of dot product scoring used with Nomic+\method{} as well as BM25 scoring used with SPLADE-v3, on selected BEIR datasets in the interest of computational efficiency.

These results appear to show that dot product scoring is unsuitable for the features extracted from an SAE as presented in our study, with strong degradation on all evaluated datasets. Interestingly, the opposite hold true for SPLADE: BM25 scoring significantly degrades performance, while dot product scoring preserves it. 

We believe these results to be expected: SPLADE is trained with an explicit regularization objective which encourages it to move away from the distributional attributes expected by BM25, while our earlier results reveal that the features extracted by our method have a term-like distribution that is particularly suitable for it, but are not shaped for inner-product scoring. 

\section{Impact of BM25 Tuning}
\label{app:tuning}

\begin{table*}[t]
\centering
\scriptsize
\setlength{\tabcolsep}{2.5pt}
\renewcommand{\arraystretch}{1.08}
\resizebox{\textwidth}{!}{%
\begin{tabular}{@{}lrrrrrrrrrrrrrrr@{}}
\toprule
Setting & Avg & SciF. & NFC & Arg. & SciD. & FiQA & TREC & Touch\'e & Quora & DBP & NQ & Clim. & Hotpot & FEVER & MSM \\
\midrule
Tuned
& 0.536
& 0.749
& 0.372
& 0.462
& 0.184
& 0.382
& 0.783
& 0.315
& 0.855
& 0.436
& 0.577
& 0.350
& 0.732
& 0.885
& 0.417
\\
Untuned
& 0.527
& 0.747
& 0.371
& 0.435
& 0.184
& 0.382
& 0.757
& 0.312
& 0.855
& 0.426
& 0.568
& 0.334
& 0.730
& 0.867
& 0.411
\\
$\Delta$
& +0.009
& +0.002
& +0.001
& +0.027
& +0.000
& +0.000
& +0.026
& +0.003
& +0.000
& +0.010
& +0.009
& +0.016
& +0.002
& +0.018
& +0.007
\\
\bottomrule
\end{tabular}%
}
\vspace{0.25em}
\caption{Comparison of tuned and untuned \method{} retrieval. All values report nDCG@10.}
\label{tab:app-bm25-tuning}
\end{table*}

Table~\ref{tab:app-bm25-tuning} shows a comparison between tuned and un-tuned results, using \method{} over Nomic. We show that tuning hyperparameters does result in a performance improvement on most datasets, but that the effect appears to be moderate, with default parameters remaining strong. Default parameters are defined as a $k1$ of 8, $b$ length penalty set to 0.7, and using square root transforms as $\phi$ for both documents and queries.

\end{document}